\documentclass[preprint,eqsecnum,aps,showpacs]{revtex4}
\usepackage{dcolumn}
\usepackage{graphicx}
\usepackage{amsmath}
\usepackage{amssymb}
\usepackage{epsfig}
\usepackage{float}
\usepackage{latexsym}
\usepackage{rotating}
\usepackage{xcolor}
\begin{document}

\title{\large {Gravitating Dyons in Vaidya Geometry}}

\vspace{5.0cm}
\author{Buddhi Vallabh Tripathi$^{1,2}${\footnote{{buddtrip@rediffmail.com}},
Hemwati Nandan$^{1,3}${\footnote{ hnandan@iucaa.ernet.in}}, Heinz Dehnen$^{3}${\footnote{heinz.dehnen@uni-konstanz.de}} and K.D. Purohit$^{4}${\footnote{kdpurohit@rediffmail.com}}}}
\vspace{0.5cm}
\affiliation{\small{
$^1$ Department of Physics, Gurukula Kangri Vishwavidyalaya, Haridwar- 249 404, India.\\
$^2$ Department of Physics, A.P.B. Govt. Post Graduate College, Agastyamuni- 246 421, India.\\
$^3$Fachbereich Physik, Universit\"{a}t Konstanz, M677,
78457 Konstanz, Germany.\\
$^4$ Department of Physics, HNB Garhwal University, Srinagar (Garhwal)- 246 174, India.}}

\vspace{2.5cm}

\begin{abstract}
\vspace{1cm}
\noindent  Gravitating monopoles and dyons in Einstein-Yang-Mills (EYM) or Einstein-Yang-Mills-Higgs (EYMH) systems have been extensively studied for various curved spacetimes, including those of black holes. We construct dyonic solutions of the EYMH theory in Vaidya spacetime using a set of generalized Julia-Zee ansatz for the fields. It is found that the dyonic charge is static in nature and it does not contribute to the energy of the \textit{null dust}.
\\

\vspace{2.0cm}
\end{abstract}
\maketitle
\newpage
\section{Introduction}
\noindent In his classic paper, Dirac, presented the argument that quantum mechanics allows for the existence of isolated magnetic poles (i.e. monopoles) and that their presence can explain the quantization of electric charge \cite{dirac}. Later, 't~Hooft \cite{hooft} and Polyakov \cite{polyakov} predicted  the existence of perfectly regular, static, finite energy, non-perturbative classical solutions of spontaneously broken non-Abelian (or Yang-Mills) gauge theories that could be interpreted as magnetic monopoles.  Julia and Zee \cite{juliazee} further extended these ideas and constructed dyonic solutions of classical non-Abelian gauge theories. It is now well--established that magnetic monopoles and dyons emerge naturally in any grand unified theory (GUT), with the electromagnetic $U(1)_{em}$ gauge group embedded in a semisimple gauge group, upon spontaneous symmetry breaking.\\

\noindent The GUT monopoles are quite massive (i.e. {\it superheavy}) to be produced in the present day accelerator experiments \cite{Drukier} such as the LHC at CERN. However, such massive objects may have been produced as topological defects in the very early universe in a symmetry breaking phase transition process \cite{Rajantie1, Rajantie2, Rajantie3}. In view of the lack of the experimental confirmation of these exotic particles, various searches (direct and indirect) at colliders, in cosmic rays and bound-in-matter for the monopoles have been made regularly \cite{Patrizii, Fairbairn, Rajantie4}.  All of these searches for monopoles are also equally sensitive to dyons. It is also worth to mention that the monopoles are also potential candidates to explain the issue of colour confinement in lattice field theory simulations (i.e. lattice  QCD) and the visualisation of monopoles therein is a subject of debate in such lattice QCD models \cite{Rajantie4, Davis}. Though the idea of monopoles and dyons is yet to be confirmed experimentally, but it is still appealing mainly due to its close connections with various crucial problems (e.g. confinement, CP violation, proton decay, cosmic strings etc.)  in particle physics, gravity and cosmology \cite{Rajantie2, Hooft2000, Confinement95, Rubakov, Lazarides, Vilenkin, Shnir}. In particular, there have been extensive studies of various aspects related to the monopoles and dyons in flat-spacetime \cite{Goddard,Preskill} .\\

\noindent In view of such diversified role of monopoles and dyons \cite{dirac}-\cite{Preskill}, the theories coupling gravity to the gauge fields namely Einstein-Yang-Mills (EYM) or Einstein-Yang-Mills-Higgs (EYMH) further led to investigations of monopoles/dyons in curved spacetime. Bais and Russell \cite{Bias} and others \cite{chofreund,perry,jutta1} presented static monopole solutions of non-Abelian (Yang-Mills) theory by solving the Einstien equations in curved spacetime. These monopole solutions in curved spacetime are also constructed with and without Bogomolny-Prasad-Sommerfeld (BPS) limit of Higgs potential \cite{ortiz}. In addition to this, the static multimonopole, monopole-antimonopole systems and the static axially symmetric multimonopole and black hole solutions in EYMH theory are also examined with greater details \cite{jutta1a, jutta2, jutta2a}. It  is also shown that such multimonopole solutions are gravitationally bound for the vanishing and small Higgs selfcoupling \cite{jutta1}.  Further the gravitating dyons (Abelian and non-Abelian) in EYMH theory with their different aspects such as large electric charge, are also investigated \cite{kamata,jutta3,jutta4}.\\

\noindent Recently, a nonstatic curved spacetime generalization of 't~Hooft-Polyakov monopole solutions in the Vaidya spacetime has been presented \cite{ghoshsingh}. Vaidya spacetime describes the gravitational field of a radiating star \cite{vaidya} and is one of the most interesting non-static solutions of the Einstein's equations in general relativity \cite{hawking,joshi}.  The nonstatic exact dyonic solutions of the Einstein-Maxwell equations in Vaidya spacetime are also studied \cite{virbhadra}. It would further be quite interesting to look for Julia-Zee type dyonic solutions for EYMH theory in Vaidya spacetime. In fact, the non-vanishing energy-momentum tensor due to the matter field may twist the spacetime considered leading to a Reissner--Nordstr\"om anti de-Sitter i.e. RN--(A)dS type solution accordingly \cite{kamata}. The issue of the stability of such solutions in view of the scaling behaviour for the mass spectrum of these solutions with respect to their charge (electric and magnetic) and independent values of the cosmological and gauge coupling constants is also an important aspect to deal with \cite{Hosotani}.\\

\noindent In the present work, we construct the dyonic solutions for EYMH theory in the Vaidya spacetime by considering a generalized class of Julia-Zee ansatz. It is found that the electric and magnetic charges remain conserved and have no contribution in the energy density of the \textit{null dust}. We examine four different possible combinations of the field parameters concerning the dyonic charge and have discussed different type of possible solutions accordingly.

\section{Lagrangian and Field Equations}
\noindent We begin with the Einstein-Yang-Mills-Higgs (EYMH) action of the form,
\begin{equation}\label{act1}
S=\int \sqrt{-g}\ d^4 x ( L_G+ L_M+{L_{ND}}),
\end{equation}
where, $L_G=R/2$ is the contribution from gravitation and $L_{ND}$ is the Lagrangian corresponding to the \textit{null dust}. The matter Lagrangian in Eq. (\ref{act1}) is given as follows
\begin{equation}\label{mL}
L_M = - \frac{1}{4} F_{\mu\nu}^a F^{\mu\nu a}-\frac{1}{2} D_\mu\Phi^a D^\mu\Phi^a - \frac{\lambda}{4}(\Phi^a \Phi^a-\xi^2)^2.
\end{equation}
 The YM field tensor $F^a_{\mu\nu}$ and the gauge-covariant derivative $D_\mu \Phi^a$ appearing in (\ref{mL}) above are defined as
\begin{eqnarray} \label{fmunu}
F^a_{\mu\nu}&=&\partial_\mu A^a_\nu - \partial_\nu A^a_\mu+ e\ f^{abc} A^b_\mu A^c_\nu\\
\label{dmu}
D_\mu \Phi^a&=&\partial_\mu \Phi^a + e\ f^{abc}A_\mu^b \Phi^c\ ,
\end{eqnarray}
where, $e$ is the gauge coupling constant, $\xi$ is the VEV of Higgs field and $f^{abc}$ are the structure constants of the gauge group. Here, we consider the gauge group SO(3) (in the adjoint representation) for which the structrue constants $f^{abc}=\epsilon^{abc}$. Variation of the action with respect to the fields ($g_{\mu\nu}, A^a_\mu$  and $\Phi^a$) yield the Einstein equations and the matter field equations, respectively as below
\begin{equation}\label{Eineqn}
R_{\mu\nu}-\frac{1}{2}g_{\mu\nu}R=T_{\mu\nu}\ ,
\end{equation}
\begin{equation}\label{mat_eqn1}
\frac{1}{\sqrt{-g}}D_\nu(\sqrt{-g}g^{\mu\rho}g^{\nu\sigma} F_{\rho\sigma}^a)= e\ \epsilon^{abc}g^{\mu\nu}(D_\nu \Phi^b)\Phi^c\ ,
\end{equation}
\begin{equation}\label{mat_eqn2}
\frac{1}{\sqrt{-g}}D_\mu(\sqrt{-g}g^{\mu\nu} D_\nu \Phi^a)=\lambda(\Phi^b \Phi^b-\xi^2) \Phi^a.
\end{equation}
The energy-momentum tensor $T_{\mu\nu}$ appearing in eqn. (\ref{Eineqn}) has contributions from the matter fields ($T_{\mu\nu}^M$) and the \textit{null dust} ($T_{\mu\nu}^{ND}$). The matter field contribution to $T_{\mu \nu}$ (i.e. $T_{\mu\nu}^M$) is given as below,
\begin{eqnarray}\label{MEMT}
T_{\mu\nu}^M&=&\left(
g^{\alpha\beta} F_{\mu\alpha}^a F_{\nu\beta}^a
+D_\mu \Phi^a D_\nu \Phi^a\right)\nonumber\\
&\ &\qquad +\ g_{\mu\nu}\left\{-\frac{1}{4}g^{\alpha\beta}g^{\gamma\delta} F_{\alpha\gamma}^a F_{\beta\delta}^a
-\frac{1}{2}g^{\alpha\beta} D_\alpha \Phi^a D_\beta \Phi^a - \frac{\lambda}{4} (\Phi^a \Phi^a-\xi^2)^2\right\}.
\end{eqnarray}
\section{Vaidya spacetime and dyonic solutions}
\noindent For our investigatation of  dyonic solutions in a Vaidya spacetime, we use the general spherically symmetric metric written in Eddington-Finklestein coordinates \cite{barrabesisrael,ghoshsingh}
\begin{equation}\label{METRIC}
ds^2=-f(v,r)\ dv^2+2\epsilon\ dv\ dr+r^2\ d\Omega^2,
\end{equation}
with $\epsilon^2=1$ and $d\Omega^2$ being the metric for the two sphere $S^2$.\\

\noindent In order to investigate the gravitating dyon solutions, we generalize the static, spherically symmetric Julia-Zee ansatz \cite{juliazee} for the gauge ($A_\mu^a$) and the Higgs field ($\Phi^a$) to the following non-static, spherically symmetric form,
\begin{equation}
A_0^a=x^a \frac{J(v,r)}{er^2}=x^a N(v,r),
\end{equation}
\begin{equation}
A_i^a=\epsilon_{aij}\ x^j \left[\frac{K(v,r)-1}{e\ r^2}\right]= \epsilon_{aij}\ x^j B(v,r),
\end{equation}
\begin{equation}
\Phi^a=x^a\left[\frac{H(v,r)}{e\ r^2}\right]=x^a\ M(v,r).
\end{equation}
For the purpose of calculations, we rewrite these ansatz in the spherical polar coordinates
\begin{equation}\label{an1}
{\bf\bar{A}_0}= r N(v,r) \widehat e_r,
\end{equation}
\begin{equation}\label{an2}
{\bf\bar{A}_r}=0\ ,\quad {\bf\bar{A}_\theta}= r^2 B(v,r)\ \widehat e_\phi\ ,\quad {\bf\bar{A}_\phi}= - r^2 B(v,r)\sin\theta\ \widehat e_\theta,
\end{equation}
\begin{equation}\label{an3}
{\bf\bar{\Phi}}= r M(v,r)\ \widehat e_r,
\end{equation}
where,
\begin{eqnarray}
\widehat e_r &\equiv&\left(\sin\theta\cos\phi,\ \sin\theta\sin\phi,\ \cos\theta\right),\\
\widehat e_\theta &\equiv&\left(\cos\theta\cos\phi,\ \cos\theta\sin\phi,\ -\sin\theta\right),\\
\widehat e_\phi &\equiv&\left(-\sin\phi,\ \cos\phi,\ 0\right),
\end{eqnarray}
are the unit-vectors in the internal space. Using the ansatz (\ref{an1})-(\ref{an3}), the components of the gauge-field tensor $F_{\mu\nu}^a$ and the gauge-covariant derivative $D_\mu \Phi^a$, simplify as below
\begin{eqnarray}\label{FandD}
{\bf\bar F_{0 r}} &=& -\left[rN\right]^{'}\ \widehat e_r,\qquad\qquad\qquad\qquad\quad
{\bf\bar F_{r \theta}} = \frac{K^{'}}{e}\ \widehat e_\phi,\nonumber\\
{\bf\bar F_{0\theta}} &=& r^2 \dot B\ \widehat e_\phi-rNK\ \widehat e_\theta,\qquad\qquad\qquad
{\bf\bar F_{r \phi}} = -\frac{K^{'}}{e}\sin\theta\ \widehat e_\theta,\nonumber\\
{\bf\bar F_{0 \phi}} &=& -(r^2 \dot B\ \widehat e_\theta+rNK\ \widehat e_\phi)\sin\theta,\qquad
{\bf\bar F_{\theta\phi}} = \frac{K^2-1}{e}\sin\theta\ \widehat e_r,\nonumber\\
D_0{\bf\bar\Phi} &=& r\dot{M}\ \widehat e_r,\qquad\qquad\qquad\qquad\qquad
D_r{\bf\bar\Phi} = \left[rM\right]^{'}\ \widehat e_r,\nonumber\\
D_\theta{\bf\bar\Phi} &=& r M K\ \widehat e_\theta, \qquad\qquad
\qquad\qquad\quad D_\phi{\bf\bar\Phi} = r M K\sin\theta\ \widehat e_\phi.
\end{eqnarray}
The {\it dot} and {\it prime} denote the partial derivatives w.r.t. $v$ and $r$, respectively, here and afterwards.\\

\noindent The electric and magnetic charges of any possible configuration are to be calculated using the 't~Hooft's gauge-invariant, generalized version of the electromagnetic tensor \cite{hooft},
\begin{equation}\label{tHtensor}
\mathcal{F}_{\mu\nu}=F_{\mu\nu}^a\ \widehat\Phi^a-\frac{1}{e}\ \epsilon_{abc}\ \widehat\Phi^a(D_\mu\widehat\Phi^b)(D_\nu\widehat\Phi^c),
\end{equation}
where $\widehat\Phi^a=\Phi^a/\sqrt{\Phi^a\Phi^a}$, while the field tensor $(F_{\mu\nu}^a)$ and the gauge covaraint derivative $(D_\nu\Phi^a)$  are given by 
Eqs (\ref{fmunu}) and  (\ref{dmu}) respectively. \\

\noindent Now using the Eqs (\ref{an3}) and (\ref{FandD}),  the non-zero components of the generalized electromagnetic tensor (\ref{tHtensor}) turn out to be
\begin{equation}\label{tHtensor2}
\mathcal{F}_{0r}=-\left[rN\right]^{'}\ ;\qquad \qquad \mathcal{F}_{\theta\phi}=\left[K^2-r^3 M^3 K^2-1\right]\frac{\sin\theta}{e}.
\end{equation}
Non-zero values of $\mathcal F_{0r}$ and $\mathcal F_{\theta\phi}$ would imply the monopole/dyon nature of the solutions, with radial electric and magnetic fields respectively.\\

\noindent Using Eqs. (\ref{MEMT}), (\ref{METRIC}), (\ref{an3}) and (\ref{FandD}), the total energy-momentum tensor acquires the following form,

\begin{eqnarray}\label{EnMoT}
T^\mu_{\ \nu}&=&\small\begin{bmatrix} 
						-[(rN){'}]^2+2\epsilon\dfrac{\dot B K^{'}}{e} & 2\epsilon\dfrac{(K^{'})^2}{r^2 e^2} + \epsilon [(rM){'}]^2 & 0 & 0\\
						+(r\dot M)(r M)^{'} & & &\\
						\\
						2\epsilon[(r\dot B)^2+N^2 K^2] & -[(rN){'}]^2+2\epsilon\dfrac{\dot B K^{'}}{e} & &\\
						+\epsilon(r\dot M)^2+2f\dfrac{\dot B K^{'}}{e} & +f[2\dfrac{(K^{'})^2}{r^2 e^2}+\{(rM)^{'}\}^2] &0 &0\\
						+f(r\dot M)(r M)^{'} +\rho(v,r)& +\ \epsilon(r\dot M)(r M)^{'} & &\\ 
						\\
						0& 0& 2\epsilon\dfrac{\dot B K^{'}}{e}+f \dfrac{(K^{'})^2}{r^2 e^2} &0\\
						& & +\dfrac{(K^2-1)^2}{r^4 e^2}+M^2 K^2 &\\
						\\
						0& 0& 0& 2\epsilon\dfrac{\dot B K^{'}}{e}+f \dfrac{(K^{'})^2}{r^2 e^2}\\
						& & & +\dfrac{(K^2-1)^2}{r^4 e^2}+M^2 K^2
				\end{bmatrix}\nonumber\\ 
&+&\delta^\mu_{\ \nu}\ \bigg[\dfrac{[(rN)^{'}]^2}{2}-2\epsilon\dfrac{\dot B K^{'}}{e} -f\dfrac{(K^{'})^2}{r^2 e^2}-\dfrac{(K^2-1)^2}{2 r^4 e^2}-\epsilon(r\dot M)(r M)^{'}\nonumber\\
&\ &\qquad\qquad-\dfrac{f}{2}[(rM)^{'}]^2-M^2 K^2-\dfrac{\lambda}{4}(r^2 M^2-\xi^2)^2\bigg],
\end{eqnarray}
where, the $\rho(v,r)$ term in the $T^0_{\ r}$ component is the \textit{null dust} contribution with $\rho(v,r)$ being the energy density of the ingoing/outgoing (pressureless) fluid \cite{ghoshsingh}.
For the metric (\ref{METRIC}), the Einstein tensor evaluates to 
\begin{eqnarray}\label{EIN_TEN}
G^\mu_{\ \nu}=\begin{bmatrix}
					\dfrac{-1+f+r f^{'}}{r^2} & 0 & 0 & 0\\\\
					\dfrac{-\dot f}{r} & \dfrac{-1+f+r f^{'}}{r^2} & 0 & 0\\\\
					0 & 0 & \dfrac{2 f^{'}+r f^{''}}{2 r} & 0\\\\
					0 & 0 & 0 & \dfrac{2 f^{'}+r f^{''}}{2 r}
		\end{bmatrix}.
\end{eqnarray}
\\
Using (\ref{EnMoT}) and (\ref{EIN_TEN}) in the Einsten equation (\ref{Eineqn}), from the $0r$ component, we obtain
\begin{eqnarray}
\left[K(v,r)\right]^{'}&=&0\ ,\\
\left[r M(v,r)\right]^{'}&=&0.
\end{eqnarray}
These equations constrain the forms of $K$ and $M$ as following,
\begin{eqnarray}
K=\beta(v)\label{aa1}\ ,\\
M=\dfrac{\alpha(v)}{r}\ ,\label{aa2}
\end{eqnarray}
where $\alpha(v)$ and $\beta(v)$ are arbitrary functions of $v$ only.
To determine $\alpha(v)$ and $\beta(v)$, we use the ansatz along with Eqs. (\ref{aa1}) and (\ref{aa2}) in the matter field equations (\ref{mat_eqn1}) and (\ref{mat_eqn2}) to obtain
 \begin{eqnarray}
\left[r^2(rN)^{'}\right]^{'}&=&0\label{mfe1}\ ,\\
r^2\dfrac{\partial}{\partial v}[(rN)^{'}]&=&2(rN)\beta^2\ ,\\
\beta(rN)^{'}&=&0\ ,\\
\beta(\beta^2-1+e r^2\alpha^2)&=&0\ ,\\
2\epsilon\dot\alpha r-\alpha[2\beta^2-\lambda r^2(\alpha^2-\xi^2)]&=&0\label{mfe5}.
 \end{eqnarray}
The various combinations of variables satisfying Eqs. (\ref{mfe1})-(\ref{mfe5}), along with the corresponding values  of the generalized electromagnetic tensor (\ref{tHtensor2}), are listed in table \ref{tab: Tb1}.
\begin{center}
\begin{table}[h]
\begin{tabular}{|c||c|c|c|c|c|c|}
\hline
{\ {\bf Case}\ } & $\ \alpha(v)\ $ & $\lambda$ & $\ \beta(v)\ $ & $N(v,r)$ & $\ \mathcal F_{0r}\ $ & $\ \mathcal F_{\theta\phi}\ $\\ \hline\hline
{\bf I} & $\pm\xi$ &\ arbitrary\ \ & & & &\\ \cline{1-3}
{\bf II} & 0 &\ arbitrary\ \  & 0 & $\ \dfrac{c_1}{r^2}+\dfrac{\gamma(v)}{r}\ $ & $\ \dfrac{c_1}{r^2}\ $ & $\ -\dfrac{\sin\theta}{e}\ $\\ \cline{1-3}
{\bf III} & $c_2$ & 0 & & & &\\ \hline
{\bf IV} & 0 &\ arbitrary\ \  & $\pm 1$ & 0 & 0 & 0\\ \hline
\end{tabular}
\caption{Various possible cases satisfying Einstein \& matter field equations. Here $c_1$, $c_2$ are arbitrary constant and $\gamma(v)$ is some arbitrary function of $v$.}\label{tab: Tb1}
\end{table}
\end{center}
\vspace{-1cm}
The Case IV as mentioned in table \ref{tab: Tb1} is trivial and the solutions carry no electric and magnetic charges. However, for the remaining cases (i.e. Cases I--III in table \ref{tab: Tb1}), the solutions carry radial electric and magnetic fields given by \cite{carroll}
\begin{eqnarray}
\mathcal E_r&=& -\mathcal F_{0r}=-\dfrac{c_1}{r^2}=\dfrac{q_e}{r^2}\ ,\label{qe}\\
\mathcal B_r&=& \dfrac{\mathcal F_{\theta\phi}}{r^2\sin\theta}=-\dfrac{1}{e r^2}=\dfrac{q_m}{r^2}.\label{qm} 
\end{eqnarray}
Here, the magnetic charge ($q_m$) is fixed by the value of the coupling constant ($e$), while the electric charge ($q_e$) is arbitrary, and the particular choice $c_1=0$ reduces these dyonic configurations to a monopole. This behaviour is similar to the behaviour of Julia-Zee dyon solutions in flat spacetime \cite{juliazee,hooft,rajaraman}. Also, the Case III with $\lambda =0$ is to be understood as the limiting case $(\lambda\rightarrow 0)$. Further, the total energy-momentum tensor (\ref{EnMoT}) for all these three cases (i.e. Cases I--III) takes the form
\begin{eqnarray}\label{EnMoT2}
T^\mu_{\ \nu}&=&\begin{bmatrix}
                               -\dfrac{({q_e}^2+{q_m}^2)}{2r^4}+\Lambda &0 &0 &0\\
				{\rho(v,r)}& -\dfrac{({q_e}^2+{q_m}^2)}{2r^4}+\Lambda &0 &0\\
				0 &0 & \dfrac{({q_e}^2+{q_m}^2)}{2r^4}+\Lambda &0\\
				0 &0 &0 & \dfrac{({q_e}^2+{q_m}^2)}{2r^4}+\Lambda.
                              \end{bmatrix}.
\end{eqnarray}
Here, $\Lambda=-\lambda(\alpha^2-\xi^2)/4$ is the contribution due to spontaneous symmetry breaking and has the values $0,\ \lambda\xi^2/4 \text{ and } 0$, for Cases I, II and III, respectively. Note, however, that the value of $\Lambda$ can be shifted by adding (or subtracting) a constant in the matter Lagrangian (\ref{mL}), which is always possible without affecting the field equations. Using Eqs. (\ref{EIN_TEN}) and (\ref{EnMoT2}) in the Einstein equation (\ref{Eineqn}), the metric-function $f(v,r)$, appearing in Eqn. (\ref{METRIC}), turns out to be \cite{ghoshsingh}
\begin{equation}\label{finalf}
 f(v,r)=1-\dfrac{2\ m(v)}{r}+\dfrac{({q_e}^2+{q_m}^2)}{2\ r^2}+\dfrac{\Lambda}{3}r^2.
\end{equation}
Here, $m(v)$, the constant of integration, is an arbitrary function of $v$. In case we switch off the electric and magnetic charges, the metric function (\ref{finalf}) reduces to that of Vaidya-Anti-de
Sitter spacetime. The energy density of the \textit{null dust} then becomes
\begin{equation}\label{finalrho}
 \rho(v,r)=-\dfrac{\dot f}{r}=2\ \dfrac{\dot m}{r^2}\,.
\end{equation}
It may be noted that in Eq.(\ref{finalf}), both the electric and magnetic charges are independent of $v$ and only the mass function is dependent on $v$. It seems that this feature itself is an artifact of the formulation used to construct these dyonic solutions. The electric and magnetic charges $(q_e\ \text{and}\ q_m)$  appearing in Eqs. (\ref{qe}) and (\ref{qm}), are not carried by the infalling/outgoing \textit{null dust} and hence are conserved unlike in case of a Vaidya-Bonnor spacetime \cite{bonnorvaidya}, where the infalling/outgoing \textit{null dust} is charged. This feature is also reflected by the absence of any charge dependent term in Eq. (\ref{finalrho}). The Kretschmann scalar which falls as $r^{-6}$ diverges along $r=0$ such that the metric with $ f(v,r)$  has scalar polynomial singular structure \cite{hawking}. In case the mass function is constant like the electric and magnetic charges then using suitable coordinate transformations, the spacetime with Eq. (\ref{finalf}) is equivalent to the spacetimes derived for non-Abelian gauge theories in curved spacetime \cite{Bias, chofreund}. 

\section{Summary, Conclusions and Future Directions}
 \noindent We have constructed the dyonic solutions of an EYMH theory in the Vaidya spactime by generalizing the Julia-Zee ansatz for the gauge and scalar fields. It is found that the charges appearing in these solutions are constant in time. In particular, the magnetic charge is given by the constant value of the gauge coupling. However, the electric charge which appears from the temporal degree of freedom of the gauge field is arbitrary.  The nature of the singularities for these solutions can be examined in detail by looking in the motion of radial null geodesics. This spacetime under different conditions reduces to the spacetimes that have already been studied \cite{Bias, chofreund}, but it would be more interesting especially to study the properties of the radiating collapsing objects \cite{joshi}.\\

\noindent Further, the construction of such dyonic solutions in higher dimensional spacetime also deserves careful attention. It would also be interesting to investigate the dyonic solutions in Vaidya-Bonnor spacetime to understand the role of charge carried by the \textit{null dust} itself. Moreover, one interesting model to look for the monopole and dyons in a Vaidya-type spacetime may be the broken symmetric theory of gravity with the Higgs field which leads to a cosmological function depending on the field excitations and it would be of crucial importance to study the behaviour of such objects with the strength of field excitations with special emphasis on the interplay between the usual cosmological constant and the cosmological function \cite{dehnen, hn, hn1, hn2}.\\

\noindent We intend to report on these issues in our future publications. \\


\noindent \textbf{Acknowledgements}\\
\noindent The authors thank Prof. S.G. Ghosh for useful discussions and Prof. B.~Kleihaus for critical comments. HN and BVT would like to thank IUCAA, Pune for support under visiting associateship program where a part of this work was done. HN is thankful to German Academic Exchange Service (DAAD), Bonn, Germany for the support under its reinvitation program.  HN is also grateful to Prof. H.~Dehnen and  Prof. Jutta Kunz for their warm hospitality at the Department of  Physics, Konstanz University and Department of Physics, Carl von Ossietzky University Oldenburg, Germany respectively. HN would also like to thank the Department of Science and Technology (DST), New Delhi for financial assistance through SR/FTP/PS-31/2009.


\begin{thebibliography}{14}
\expandafter\ifx\csname natexlab\endcsname\relax\def\natexlab#1{#1}\fi
\expandafter\ifx\csname bibnamefont\endcsname\relax
  \def\bibnamefont#1{#1}\fi
\expandafter\ifx\csname bibfnamefont\endcsname\relax
  \def\bibfnamefont#1{#1}\fi
\expandafter\ifx\csname citenamefont\endcsname\relax
  \def\citenamefont#1{#1}\fi
\expandafter\ifx\csname url\endcsname\relax
  \def\url#1{\texttt{#1}}\fi
\expandafter\ifx\csname urlprefix\endcsname\relax\def\urlprefix{URL }\fi
\providecommand{\bibinfo}[2]{#2}
\providecommand{\eprint}[2][]{\url{#2}}


\bibitem{dirac}
\bibinfo{author}{\bibfnamefont{Dirac}, \bibnamefont{P.A.M.}}:
\bibinfo{title}{Quantised singularities in the electromagnetic field}.
  \bibinfo{journal}{Proc. Roy. Soc. Lond. A} \textbf{\bibinfo{volume}{133}},
  \bibinfo{pages}{60} (\bibinfo{year}{1931}).

\bibitem{hooft}
\bibinfo{author}{\bibfnamefont{'t Hooft}, \bibnamefont{G.}}:
\bibinfo{title}{Magnetic monopoles in unified gauge theories}.
  \bibinfo{journal}{Nucl. Phys. B} \textbf{\bibinfo{volume}{79}},
  \bibinfo{pages}{276} (\bibinfo{year}{1974}).

\bibitem{polyakov}
\bibinfo{author}{\bibfnamefont{Polyakov}, \bibnamefont{A.M.}}:
\bibinfo{title}{Particle spectrum in quantum field theory}.
  \bibinfo{journal}{JETP Lett.} \textbf{\bibinfo{volume}{20}},
  \bibinfo{pages}{194} (\bibinfo{year}{1974}) [Pis'ma Zh. Eksp. Teor. Fiz. \textbf{20}, 430 (1974)].

\bibitem{juliazee}
\bibinfo{author}{\bibfnamefont{Julia}, \bibnamefont{B.}},
 \bibinfo{author}{\bibfnamefont{Zee}, \bibnamefont{A.}}:
\bibinfo{title}{Poles with both magnetic and electric charges in non-Abelian gauge theory}.
  \bibinfo{journal}{Phys. Rev. D} \textbf{\bibinfo{volume}{11}},
  \bibinfo{pages}{2227} (\bibinfo{year}{1975}).

\bibitem{Drukier}
\bibinfo{author}{\bibfnamefont{Drukier}, \bibnamefont{A.}},
 \bibinfo{author}{\bibfnamefont{Nussinov}, \bibnamefont{S.}}:
\bibinfo{title}{Monopole pair creation in energetic collisions: Is it possible?} \bibinfo{journal} {Phys. Rev. Lett.} \textbf{\bibinfo{volume}{49}},
  \bibinfo{pages}{102} (\bibinfo{year}{1982}).

\bibitem{Rajantie1}
\bibinfo{author}{\bibfnamefont{Rajantie}, \bibnamefont{A.}}:
\bibinfo{title}{Magnetic monopoles from gauge theory phase transitions.} 
\bibinfo{journal} {Phys. Rev. D} \textbf{\bibinfo{volume}{68}},
  \bibinfo{pages}{021301} (\bibinfo{year}{2003}).

\bibitem{Rajantie2}
\bibinfo{author}{\bibfnamefont{Rajantie}, \bibnamefont{A.}}:
\bibinfo{title}{Defect formation in the early universe.} \bibinfo{journal} {Contemp. Phys.} \textbf{\bibinfo{volume}{44}},
  \bibinfo{pages}{485} (\bibinfo{year}{2003}).

\bibitem{Rajantie3}
\bibinfo{author}{\bibfnamefont{Rajantie}, \bibnamefont{A.}}:
\bibinfo{title}{Formation of topological defects in gauge field theories.} \bibinfo{journal} {Int. J. Mod. Phys. A} \textbf{\bibinfo{volume}{17}},
  \bibinfo{pages}{1} (\bibinfo{year}{2002}).

\bibitem{Patrizii}
\bibinfo{author}{\bibfnamefont{Giacomelli}, \bibnamefont{G.}},
 \bibinfo{author}{\bibfnamefont{Patrizii}, \bibnamefont{L.}}:
\bibinfo{title}{Magnetic monopole searches.} \textbf{\bibinfo{volume}{arXiv:hep-ex/0506014}}.


\bibitem{Fairbairn}
\bibinfo{author}{\bibfnamefont{Fairbairn}, \bibnamefont{M. et al.}}:
\bibinfo{title}{Stable massive particles at colliders.} \bibinfo{journal} {Phys. Rept.}  \textbf{\bibinfo{volume}{438}},
  \bibinfo{pages}{1} (\bibinfo{year}{2007}).

\bibitem{Rajantie4}
\bibinfo{author}{\bibfnamefont{Rajantie}, \bibnamefont{A.}}:
\bibinfo{title}{Mass of a quantum 't Hooft-Polyakov monopole.}  \bibinfo{journal} {JHEP} \textbf{\bibinfo{volume}{0601}},
  \bibinfo{pages}{088} (\bibinfo{year}{2006}).

\bibitem{Davis}
\bibinfo{author}{\bibfnamefont{Davis}, \bibnamefont{A.}},
 \bibinfo{author}{\bibfnamefont{Kibble}, \bibnamefont{T.}},
 \bibinfo{author}{\bibfnamefont{Rajantie}, \bibnamefont{A.}}, 
\bibinfo{author}{\bibfnamefont{Shanahan}, \bibnamefont{H.}}:
\bibinfo{title}{Topological defects in lattice gauge theories.} \bibinfo{journal} {JHEP} \textbf{\bibinfo{volume}{0011}}, 
  \bibinfo{pages}{010} (\bibinfo{year}{2000}).


\bibitem{Hooft2000}
\bibinfo{author}{\bibfnamefont{'t Hooft}, \bibnamefont{G.}},
 \bibinfo{author}{\bibfnamefont{Bruckmann}, \bibnamefont{F.}}:
\bibinfo{title}{Monopoles, instantons and confinement. \textbf
{arXiv:hep-th/0010225}}.

\bibitem{Confinement95}
\bibinfo{author}{\bibfnamefont{Toki}, \bibnamefont{H. et al.}}:
\bibinfo{title}{Proceedings of International RCNP Workshop on Colour Confinement and Hadrons (Confinement 95), World Scientific Publishing Co.
Pvt. Ltd., Singapore (1995)}.


\bibitem{Rubakov}
\bibinfo{author}{\bibfnamefont{Rubakov}, \bibnamefont{V.A.}}:
\bibinfo{title}{ Superheavy magnetic monopoles and proton decay.} \bibinfo{journal} {JETP Lett. } \textbf{\bibinfo{volume}{33}},
  \bibinfo{pages}{644} (\bibinfo{year}{1981}).

\bibitem{Lazarides}
\bibinfo{author}{\bibfnamefont{Lazarides}, \bibnamefont{G.}},
 \bibinfo{author}{\bibfnamefont{Shafi}, \bibnamefont{Q.}},
\bibinfo{Walsh}{\bibfnamefont{Walsh}, \bibnamefont{T.F.}}:
\bibinfo{title}{Cosmic strings and domains in unified theories.} \bibinfo{journal} { Nucl. Phys. B } \textbf{\bibinfo{volume}{195}}, 
  \bibinfo{pages}{157} (\bibinfo{year}{1982}).

\bibitem{Vilenkin}
\bibinfo{author}{\bibfnamefont{Vilenkin}, \bibnamefont{J.}}:
\bibinfo{title}{Cosmological evolution of monopoles connected by strings.} \bibinfo{journal} {Nucl. Phys. B}  \textbf{\bibinfo{volume}{240}}, 
  \bibinfo{pages}{196} (\bibinfo{year}{1982}).

\bibitem{Shnir}
\bibinfo{author}{\bibfnamefont{Shnir}, \bibnamefont{Y.M.}}:
\emph{\bibinfo{title}{Magnetic monopoles}} ( Springer, Berlin, 2005).
 
\bibitem{Goddard}
\bibinfo{author}{\bibfnamefont{Goddard}, \bibnamefont{P.}},
 \bibinfo{author}{\bibfnamefont{Olive}, \bibnamefont{D.I.}}:
\bibinfo{title}{Magnetic monopoles in gauge field theories.} \bibinfo{journal} { Rep. Prog. Phys.} \textbf{\bibinfo{volume}{41}}, 
  \bibinfo{pages}{1357} (\bibinfo{year}{1978}).

\bibitem{Preskill}
\bibinfo{author}{\bibfnamefont{Preskill}, \bibnamefont{J.}}:
\bibinfo{title}{ Magnetic monopoles.} \bibinfo{journal} {Ann. Rev. Nucl. Part. Sci.}  \textbf{\bibinfo{volume}{34}}, 
  \bibinfo{pages}{461} (\bibinfo{year}{1984}).

\bibitem{Bias} 
\bibinfo{author}{\bibfnamefont{Bais}, \bibnamefont{F.A.}},
 \bibinfo{author}{\bibfnamefont{Russell}, \bibnamefont{R.J.}}:
\bibinfo{title}{ Magnetic-monopole solution of non-Abelian gauge theory in curved spacetime.} \bibinfo{journal} {Phys. Rev. D}  \textbf{\bibinfo{volume}{11}}, 
  \bibinfo{pages}{2692} (\bibinfo{year}{1975}).

\bibitem{chofreund}
\bibinfo{author}{\bibfnamefont{Cho}, \bibnamefont{Y.M.}},
 \bibinfo{author}{\bibfnamefont{Freund}, \bibnamefont{P.G.O.}}:
\bibinfo{title}{Gravitating 't~Hooft monopoles}.
  \bibinfo{journal}{Phys. Rev. D} \textbf{\bibinfo{volume}{12}},
  \bibinfo{pages}{1588} (\bibinfo{year}{1975}).

\bibitem{perry}
\bibinfo{author}{\bibfnamefont{Nieuwenhuizen}, \bibnamefont{P. van}},
\bibinfo{author}{\bibfnamefont{Wilkinson}, \bibnamefont{D.}},
\bibinfo{author}{\bibfnamefont{Perry}, \bibnamefont{M.J.}}:
\bibinfo{title}{Regular solution of 't~Hooft magnetic monopole model in curved space.}
  \bibinfo{journal}{Phys. Rev. D} \textbf{\bibinfo{volume}{13}},
  \bibinfo{pages}{778} (\bibinfo{year}{1976}).

\bibitem{jutta1}
\bibinfo{author}{\bibfnamefont{Hartmann}, \bibnamefont{B.}},
\bibinfo{author}{\bibfnamefont{Kleihaus}, \bibnamefont{B.}},
\bibinfo{author}{\bibfnamefont{Jutta}, \bibnamefont{K.}}:
\bibinfo{title}{Gravitationally bound monopoles}.
  \bibinfo{journal}{Phys. Rev. Lett.} \textbf{\bibinfo{volume}{86}},
  \bibinfo{pages}{1422} (\bibinfo{year}{2001}).

\bibitem{ortiz}
\bibinfo{author}{\bibfnamefont{Ortiz}, \bibnamefont{M.E.}}:
\bibinfo{title}{Curved-space magnetic monopoles}.
  \bibinfo{journal}{Phys. Rev. D} \textbf{\bibinfo{volume}{45}},
  \bibinfo{pages}{R2586} (\bibinfo{year}{1992}).

\bibitem{jutta1a}
\bibinfo{author}{\bibfnamefont{Kleihaus}, \bibnamefont{B.}},
\bibinfo{author}{\bibfnamefont{Kunz}, \bibnamefont{J.}}:
\bibinfo{title}{ 	
Monopole - anti-monopole solutions of Einstein-Yang-Mills-Higgs theory}.
  \bibinfo{journal}{Phys. Rev. Lett.} \textbf{\bibinfo{volume}{85}},
  \bibinfo{pages}{2430} (\bibinfo{year}{2000}).

\bibitem{jutta2}
\bibinfo{author}{\bibfnamefont{Kleihaus}, \bibnamefont{B.}},
\bibinfo{author}{\bibfnamefont{Kunz}, \bibnamefont{J.}},
\bibinfo{author}{\bibfnamefont{Shnir}, \bibnamefont{Y.}}:
\bibinfo{title}{Gravitating monopole-antimonopole chains and vortex rings}.
  \bibinfo{journal}{Phys. Rev. D} \textbf{\bibinfo{volume}{71}},
  \bibinfo{pages}{024013} (\bibinfo{year}{2005}).

\bibitem{jutta2a}
\bibinfo{author}{\bibfnamefont{Hartmann}, \bibnamefont{B.}},
\bibinfo{author}{\bibfnamefont{Kleihaus}, \bibnamefont{B.}},
\bibinfo{author}{\bibfnamefont{Kunz}, \bibnamefont{J.}}: Axially symmetric monopoles and black holes in Einstein-Yang-Mills-Higgs theory. \bibinfo{title}{}
  \bibinfo{journal}{Phys. Rev. D} \textbf{\bibinfo{volume}{65}},
  \bibinfo{pages}{024027} (\bibinfo{year}{2002}).

\bibitem{kamata}
\bibinfo{author}{\bibfnamefont{Kamata}, \bibnamefont{M.}}:
\bibinfo{title}{Abelian and non-Abelian dyon solutions in curved space-time}.
  \bibinfo{journal}{Prog. Theo. Phys.} \textbf{\bibinfo{volume}{68}},
  \bibinfo{pages}{960} (\bibinfo{year}{1982}).

\bibitem{jutta3}
\bibinfo{author}{\bibfnamefont{Ibadov}, \bibnamefont{R.}},
\bibinfo{author}{\bibfnamefont{Kleihaus}, \bibnamefont{B.}},
\bibinfo{author}{\bibfnamefont{Kunz}, \bibnamefont{J.}},
\bibinfo{author}{\bibfnamefont{Neemann}, \bibnamefont{U.}}:
\bibinfo{title}{Gravitating dyons with large electric charge}.
  \bibinfo{journal}{Phys. Lett. B} \textbf{\bibinfo{volume}{659}},
  \bibinfo{pages}{421} (\bibinfo{year}{2008}).

\bibitem{jutta4}
\bibinfo{author}{\bibfnamefont{Kleihaus}, \bibnamefont{B.}},
\bibinfo{author}{\bibfnamefont{Kunz}, \bibnamefont{J.}},
\bibinfo{author}{\bibfnamefont{Navarro-Lerida}, \bibnamefont{F.}},
\bibinfo{author}{\bibfnamefont{Neemann}, \bibnamefont{U.}}:
\bibinfo{title}{Stationary dyonic regular and black hole solutions}.
  \bibinfo{journal}{Gen. Rel. Grav.} \textbf{\bibinfo{volume}{40}},
  \bibinfo{pages}{1279} (\bibinfo{year}{2008}).

\bibitem{ghoshsingh}
\bibinfo{author}{\bibfnamefont{Ghosh}, \bibnamefont{S.G.}},
 \bibinfo{author}{\bibfnamefont{Singh}, \bibnamefont{L.P.}}:
\bibinfo{title}{Gravitating magnetic monopole in Vaidya geometry}.
  \bibinfo{journal}{Phys. Rev. D} \textbf{\bibinfo{volume}{83}},
  \bibinfo{pages}{067501} (\bibinfo{year}{2011}).


\bibitem{vaidya}
\bibinfo{author}{\bibfnamefont{Vaidya}, \bibnamefont{P.C.}}:
\bibinfo{title}{The external field of a radiating star in general relativity}.
  \bibinfo{journal}{Curr. Sci.} \textbf{\bibinfo{volume}{12}},
  \bibinfo{pages}{183} (\bibinfo{year}{1943}).
[Reprinted in {Gen. Rel. Grav.} {\bf 31}, 119 (1999)].


\bibitem{hawking}
\bibinfo{author}{\bibfnamefont{Hawking}, \bibnamefont{S.W.}},
\bibinfo{author}{\bibfnamefont{Ellis}, \bibnamefont{G.F.R.}}:
\emph{\bibinfo{title}{The large scale structure of spacetime}}
  \bibinfo{journal}{(Cambridge University Press, Cambridge, UK, 1973)}.

\bibitem{joshi}
\bibinfo{author}{\bibfnamefont{Joshi}, \bibnamefont{P.S.}}:
\emph{\bibinfo{title}{Global aspects in gravitation and cosmology}}
  \bibinfo{journal}{(Oxford University Press, Oxford, UK, 1997)}.


\bibitem{virbhadra}
\bibinfo{author}{\bibfnamefont{Chamorro}, \bibnamefont{A.}},
 \bibinfo{author}{\bibfnamefont{Virbhadra}, \bibnamefont{K.S.}}:
\bibinfo{title}{A radiating dyon solution}.
  \bibinfo{journal}{Pramana} \textbf{\bibinfo{volume}{45(2)}},
  \bibinfo{pages}{181} (\bibinfo{year}{1995}).

\bibitem{barrabesisrael}
\bibinfo{author}{\bibfnamefont{Barrabes}, \bibnamefont{C.}},
 \bibinfo{author}{\bibfnamefont{Israel}, \bibnamefont{W.}}:
\bibinfo{title}{Thin shells in general relativity and cosmology: The lightlike limit}.
  \bibinfo{journal}{Phys. Rev. D} \textbf{\bibinfo{volume}{43}},
  \bibinfo{pages}{1129} (\bibinfo{year}{1991}).


\bibitem{Hosotani}
\bibinfo{author}{\bibfnamefont{Hosotani}, \bibnamefont{Y.}},
\bibinfo{title}{Scaling behavior in the Einstein-Yang-Mills monopoles
and dyons}.
  \bibinfo{journal}{J Math. Phys.} \textbf{\bibinfo{volume}{43}},
  \bibinfo{pages}{597} (\bibinfo{year}{2002}).



\bibitem{carroll}
\bibinfo{author}{\bibfnamefont{Carroll}, \bibnamefont{S.}}:
  \emph{\bibinfo{title}{Spacetime and Geometry: An introduction to general relativity}}
  (\bibinfo{publisher}{Addison Wesley, USA},
  \bibinfo{year}{2004}) [pp. 254-255].

\bibitem{rajaraman}
\bibinfo{author}{\bibfnamefont{Rajaraman}, \bibnamefont{R.}}:
  \emph{\bibinfo{title}{Solitons and Instantons}}
  (\bibinfo{publisher}{North-Holland Publishing Company},
  \bibinfo{year}{1982}).

\bibitem{bonnorvaidya}
\bibinfo{author}{\bibfnamefont{Bonnor}, \bibnamefont{W.B.}},
 \bibinfo{author}{\bibfnamefont{Vaidya}, \bibnamefont{P.C.}}:
\bibinfo{title}{Spherically symmetric radiation of charge in Einstein-Maxwell Theory}.
  \bibinfo{journal}{Gen. Rel. Grav.} \textbf{\bibinfo{volume}{1}},
  \bibinfo{pages}{127} (\bibinfo{year}{1970}).

\bibitem{dehnen}
\bibinfo{author}{\bibfnamefont{Dehnen}, \bibnamefont{H.}},
 \bibinfo{author}{\bibfnamefont{Frommert}, \bibnamefont{H.}},
 \bibinfo{author}{\bibfnamefont{Ghaboussi}, \bibnamefont{F.}}:
\bibinfo{title}{Higgs field and a new scalar-tensor theory of gravity.} 
\bibinfo{journal}{{Int. J. Theor. Phys.}} \textbf{\bibinfo{volume}{31}},
 \bibinfo{pages}{109} (\bibinfo{year}{1992}).

\bibitem{hn}
 \bibinfo{author}{\bibfnamefont{Bezares-Roder}, \bibnamefont{Nils M.}},
\bibinfo{author}{\bibfnamefont{Nandan}, \bibnamefont{H.}},
\bibinfo{author}{\bibfnamefont{Dehnen}, \bibnamefont{H.}}:
\bibinfo{title}{Horizon-less spherically symmetric vacuum-solutions in a Higgs scalar-tensor theory of gravity.} \bibinfo{journal}{Int. J. Theor. Phys.} \textbf{\bibinfo{volume}{46}},
 \bibinfo{pages}{2429} (\bibinfo{year}{2007}).

\bibitem{hn1}
\bibinfo{author}{\bibfnamefont{Nandan}, \bibnamefont{H.}},
 \bibinfo{author}{\bibfnamefont{Bezares-Roder}, \bibnamefont{Nils M.}},
 \bibinfo{author}{\bibfnamefont{Dehnen}, \bibnamefont{H.}}:
\bibinfo{title}{ 	
Black hole solutions and pressure terms in induced gravity with Higgs potential.} \bibinfo{journal}{Class. Quant. Grav.} \textbf{\bibinfo{volume}{27}},
 \bibinfo{pages}{245003} (\bibinfo{year}{2010}).
\bibitem{hn2}
 \bibinfo{author}{\bibfnamefont{Bezares-Roder}, \bibnamefont{Nils M.}},
\bibinfo{author}{\bibfnamefont{Nandan}, \bibnamefont{H.}},
 \bibinfo{author}{\bibfnamefont{Dehnen}, \bibnamefont{H.}}:
\bibinfo{title}{ 	Scalar-field pressure in induced gravity with Higgs potential and dark matter.}  \bibinfo{journal}{JHEP} \textbf{\bibinfo{volume}{1010}}, \bibinfo{pages}{113} (\bibinfo{year}{2010}).





\end{thebibliography}
\end{document}